\documentclass[12pt]{article}
\usepackage{amssymb,amsfonts,amsmath,amsthm}
\usepackage[dvips]{graphicx}
\newcommand{\mesp}{(\Omega, {\mathcal B})}
\newcommand{\half}{\frac{1}{2}}

\newcommand{\ra}{\rightarrow}
\newcommand{\Ra}{\Rightarrow}
\newcommand{\witi}{\widetilde}

\newcommand{\nn}{{\mathbb N}}

\newcommand{\calp}{{\mathcal P}}
\newcommand{\calo}{{\mathcal O}}
\newcommand{\call}{{\mathcal L}}

\newcommand{\calm}{{\mathcal M}}
\newcommand{\caln}{{\mathcal N}}
\newcommand{\calf}{{\mathcal F}}
\newcommand{\cale}{{\mathcal E}}
\newcommand{\calb}{{\mathcal B}}
\newcommand{\veps}{\varepsilon}
\newcommand{\beq}{\begin{eqnarray*}}
\newcommand{\feq}{\end{eqnarray*}}
\newcommand{\beqn}{\begin{eqnarray}}
\newcommand{\feqn}{\end{eqnarray}}
\newcommand{\be}{\begin{equation*}}
\newcommand{\fe}{\end{equation*}}
\newcommand{\ben}{\begin{equation}}
\newcommand{\fen}{\end{equation}}

\makeatletter \@addtoreset{equation}{section} \makeatother

\newcounter{vadik}
\newtheorem{theorem}{Theorem}
\newtheorem{definition}[equation]{Definition}
\newtheorem{lemma}[equation]{Lemma}

\newtheorem*{theorem*}{Theorem}
\newtheorem{proposition}[equation]{Proposition}

\newtheorem*{corollary*}{Corollary}

\newtheorem*{remark*}{Remark}

\newtheorem{example}[equation]{Example}

\newtheorem{itexercise}{Exercise}

\newtheorem*{itexdif8}{*Exercise 8}

\newtheorem*{itexdif9}{*Exercise 9}

\newtheorem*{itexdif10}{*Exercise 10}

\newtheorem*{ithint}{Hint}

\newtheorem*{itinstruction}{Instruction}

\newtheorem{itsolution1}{Solution}

\newtheorem*{itsolution}{Solution}

\title{On Probabilistic Analog Automata} \author{ Asa
  Ben-Hur\thanks{Department of Biochemistry, B400 Beckman Center,
    Stanford University, CA 94305-5307, USA.} \and Alexander
  Roitershtein\footnote{Department of
    Mathematics, Technion - IIT, Haifa 32000, Israel (e-mail:
    roiterst@tx.technion.ac.il). } \and Hava T.~
  Siegelmann\footnote{Department of Computer Science, University of
    Massachusetts at Amherst, 710 N. Pleasant Street
    Amherst, MA 01003-9305 USA.} \and \\
$$}
\begin{document}
\maketitle
\begin{abstract} We consider probabilistic automata on a general 
  state space and study their computational power. The model is based
  on the concept of language recognition by probabilistic automata due
  to Rabin \cite {rabin} and models of analog computation in a noisy
  environment suggested by Maass and Orponen \cite{orponen}, and Maass
  and Sontag \cite{sontag}. Our main result is a generalization of
  Rabin's reduction theorem that implies that under very mild
  conditions, the computational power of the automaton is limited to
  regular languages.
\end{abstract}
\section{Introduction}
Probabilistic automata have been studied since the early 60's
\cite{paz}. Relevant to our line of interest is the work of Rabin
\cite{rabin} where probabilistic (finite) automata with isolated
cut-point were introduced. He showed that such automata recognize
regular languages, and identified a condition which restricts them to
definite languages (languages for which there exists an integer $r$
such that any two words coinciding on the last $r$ symbols are both or
neither in the language).

Paz generalized Rabin's condition for definite languages and
called it {\em weak ergodicity}. He showed that Rabin's stability
theorem holds for weakly ergodic systems as well
\cite{paz-aw,paz}. 

In recent years there is much interest in analog automata and their
computational properties. A model of analog computation in a noisy
environment was introduced by Maass and Orponen in \cite{orponen}.
For a specific type of noise it recognizes only regular languages (see
also \cite{casey}).  Analog neural networks with Gaussian-like noise
were shown by Maass and Sontag \cite{sontag} to be limited in their
language-recognition power to definite languages. This is in sharp
contrast with the noise-free case where analog computational models
are capable of simulating Turing machines, and when containing real
constants, can recognize non-recursive languages \cite{BOOK}.
 
In this work we propose a model which includes the discrete model of
Rabin and the analog models suggested in \cite{orponen,sontag}, and
find general conditions (related to ergodic properties of stochastic
kernels representing probabilistic transitions of the automaton) that
restrict its computational power to regular and definite languages.

We denote the state space of the automaton by $\Omega$ and the
alphabet by $\Sigma$. As usual, the set of all words of length $r$ is
denoted by $\Sigma^r$ and $\Sigma^*:=\cup_{r \in \nn} \Sigma^r.$ We
assume that $\Omega$ is a Polish space and denote by $\calb$ the
$\sigma$-algebra of its Borel subsets.

Let $\mathcal E$ be the Banach space of signed measures on
$(\Omega,{\mathcal B})$ with the total variation norm
$$\|\mu\|_1:= \sup_{A \in \calb}\mu(A)-\inf_{A \in \calb }\mu(A),$$
and let $\call$ be the space of bounded linear operators in $\mathcal
E$ with the norm $\|P\|_1= \sup\limits_{\|\mu\|_1=1} \|P\mu\|_1$.
\begin{definition}
\label{mar-oper}
An operator $P \in \call$ is said to be a {\em Markov operator}
if for any probability measure $\mu$, the image $P\mu$ is again a
probability measure. A {\em Markov system} is a set of Markov
operators $T=\{P_u: u \in \Sigma\}.$
\end{definition}
With any Markov system $T$, one can associate a probabilistic
computational system as follows.  At each computation step the system
receives an input signal $u \in \Sigma$ and updates its state.  If the
probability distribution on the initial states is given by the
probability measure $\mu_{0} ,$ then the distribution of states after
$n+1$ computational steps on inputs $w=w_{0},w_{1},...,w_{n},$ is
defined by $$P_{w}\mu_{0}=P_{w_{n}}\cdot \ldots \cdot P_{w_{1}}
P_{w_{0}}\mu_{0}.$$
If the probability of moving from state $x \in
\Omega$ to set $A \in \calb$ upon receiving input $u \in \Sigma$ is
given by a stochastic kernel $P_u(x,A),$ then
$P_u\mu(A)=\int_{\Omega}P_u(x,A)\mu(dx).$

Let $\mathcal A$ and $\mathcal R$ be two subsets of $\calp$ with the
property of having a $\rho$-gap
\begin{equation}
\label{dist}
\mbox{dist}({\mathcal A}, {\mathcal R})= \inf_{\mu \in {\mathcal A}, \nu \in
{\mathcal R}}
\|\mu-\nu\|_1= \rho >0
\end{equation}
A Markov computational system becomes a language recognition device by
agreement that an input string is accepted or rejected according to
whether the distribution of states of the MCS after reading the string
is in $\mathcal A$ or in $\mathcal R$.

Finally, we have the definition:
\begin{definition}
\label{drecognition}
Let $\mu_{0}$ be an initial distribution and $\mathcal A$ and
$\mathcal R$ be two bounded subsets of $\cale$ that satisfy
$(\ref{dist}).$ Let $T=\{P_u:u\in \Sigma\}$ be a set of Markov
operators on $\cale$.  We say that the {\em Markov computational
  system} (MCS) $\calm=\langle \cale,$ ${\mathcal A},$${\mathcal R},
\Sigma, \mu_0, T \rangle $ {\em recognizes} the subset $L\subseteq
\Sigma^*$ if for all $w \in \Sigma^*$: $$w \in L \Leftrightarrow
P_w\mu_0\in {\mathcal A}$$
$$~w \notin L \Leftrightarrow P_w\mu_0\in {\mathcal R}.$$
\end{definition}
We recall that two words $u,v \in \Sigma^{*}$ are equivalent with
respect to $L$ if and only if $uw \in L \Leftrightarrow vw \in L$ for
all $w \in \Sigma^*.$ A language $L \subseteq \Sigma^*$ is {\em regular} if
there are finitely many equivalence classes. $L$ is {\em definite} if for
some $r>0,$ $wu \in L \Leftrightarrow u \in L$ for all $w \in
\Sigma^{*}$ and $u \in \Sigma^r.$ If $\Sigma$ is finite, then definite
languages are regular.

A quasi-compact MCS can be characterized as a system such that
$\Sigma$ is finite and there is a set of compact operators $\{Q_w \in
\call : w \in \Sigma^*\}$ such that $\lim_{|w| \ra
  \infty}\|P_w-Q_w\|_1=0.$ Section \ref{red-lemma} is devoted to MCS
having this property. Our main result (Theorem \ref{yosida239}) states
that quasi-compact MCS can recognize regular languages only. As a
consequence of this result, we obtain the following theorem which shows
that ``any reasonable'' probabilistic automata recognize regular
languages only:
\begin{theorem*}
  Let $\calm$ be an MCS. Assume that $\Sigma$ is finite, and there
  exist constant $K>0$ and probability measure $\mu$ such that
  $P_u(x,A) \leq K \mu(A)$ for all $u \in \Sigma,$ $x \in \Omega,$ $A
  \in \calb.$ Then, if a language $L \subseteq \Sigma^{*}$ is
  recognized by $\mathcal M$, it is a regular language.
\end{theorem*}
A MCS is weakly ergodic if there is a set of constant operators $\{H_w
\in \call : w \in \Sigma^*\}$ such that $\lim_{|w| \ra
  \infty}\|P_w-H_w\|_1=0.$ In Section \ref{ewak} we carry over the
theory of discrete weakly ergodic systems developed by Paz
\cite{paz-aw, paz} to our general setup. In particular, if a language
$L$ is recognized by a weakly ergodic MCS, then it is definite
language.
\section{The Reduction Lemma and Quasi-compact MCS}
\label{red-lemma} We prove here a general version of Rabin's
reduction theorem (Lemma \ref{up-bound}) which makes the connection
between a measure of non-compactness of the set $\{P_w\mu_0:w \in
\Sigma^*\}$ with the computational power of MCS. Then we introduce the
notion of quasi-compact MCS and show that these systems satisfy the
conditions stated in Lemma \ref{up-bound}.

If $S$ is a bounded subset of a Banach space $E,$ Kuratowski's measure
of non-compactness $\alpha(S)$ of $S$ is defined by \cite{banas} \beqn
\alpha(S)&=&\inf\{\veps>0 : S~\mbox{can be covered by a finite number
  of sets} \nonumber \\&&~~~~~ \mbox{of diameter smaller than}~
\veps\}. \feqn A bounded set $S$ is totally bounded if $\alpha(S)=0.$
\begin{lemma}
\label{up-bound} Let $\mathcal M$ be an MCS, and assume that
$\alpha(\calo)<\rho$, where $\calo=\{P_w\mu_0:w \in \Sigma^*\}$ is
the set of all possible state distributions of $\calm,$ and
$\rho$ is defined by \eqref{dist}. Then, if a language $L
\subseteq \Sigma^{*}$ is recognized by $\mathcal M$, it is a
regular language.
\end{lemma}
\begin{proof}
If $\|P_{u}\mu_{0}-P_{v}\mu_{0} \|_{1} < \rho,$ then $u$ and $v$
are in the same equivalence class. Indeed, for any $w \in
\Sigma^*,$ \beq \|P_{uw}\mu_0-P_{vw}\mu_0 \|_1 =
\|P_w\left(P_u\mu_0-P_v\mu_0\right)\|_1 \leq \|P_u\mu_0-P_v\mu_0
\|_1 < \rho.\feq There is at most a finite number of equivalence
classes, since there is a finite covering of $\mathcal O$ by sets
with diameter less than $\rho$.
\end{proof}
Lemma \ref{up-bound} is a natural generalization of Rabin's
reduction theorem \cite{rabin}, where the state space $\Omega$ is
finite, and hence the whole space of probability measures is
compact.
\begin{example}
  Consider an MCS $\calm$ such that $\Omega=\nn$ and $\Sigma$ is a
  finite set. If the sums $\Sigma_{j}P_{u}(i,j)$ converges uniformly
  for each $u \in \Sigma$, then the corresponding operators $P_u \in
  \mathcal L$ are compact \cite{cohen-dun}, and consequently (since
  $\calo \subset \cup_{u \in \Sigma} P_u \calp$) $\calm$ recognizes
  regular languages only.
\end{example}
Recall that a Markov operator $P$ is called quasi-compact if there is
a compact operator $Q \in \call$ such that $\|P-Q\|_1 <1$
\cite{neveu}.
\begin{definition}
\label{qc} An MCS $\mathcal M$ is called quasi-compact if the alphabet 
$\Sigma$ is finite, and there exist constants $r,\delta >0$ such that
for any $w \in \Sigma^r$ there is a compact operator $Q_w$ which
satisfies $\|P_w-Q_w\|_1 \leq 1-\delta.$
\end{definition}
If an MCS $\calm$ is quasi-compact, then there exists a constant $M>0$
and a collection of compact operators $\{Q_w:w \in \Sigma^*\}$ such
that $\|P_w-Q_w\|_1 \leq M(1-\delta)^{|w|/r},$ for all $w \in \Sigma^*.$

The next theorem characterizes the computational power of
quasi-compact MCS.
\begin{theorem}
\label{yosida239} If $\mathcal M$ is a quasi-compact MCS, and a
language $L \subseteq \Sigma^{*}$ is recognized by $\mathcal M$,
then it is a regular language.
\end{theorem}
\begin{proof}
Fix any $\veps>0.$ There exist a number $n \in \nn$ and compact
operators $Q_w, ~w \in \Sigma^n$ such that $\|P_w-Q_w\|_1 \leq
\veps$ for all $w \in \Sigma^n.$ For any words $v \in \Sigma^{*}$
and $w \in \Sigma^{n}$, we have $\|P_{vw}\mu_0-Q_w(P_v\mu_0)\|_1
\leq \|P_w-Q_w\|_1 \leq \veps$. Since
$Q_{w}\left(P_{v}\mu_{0}\right)$ is an element of the totally
bounded set $Q_{w}\left ( {\mathcal P} \right)$, then the last
inequality implies that the set ${\mathcal O}=\{P_{u}\mu_{0}: u
\in \Sigma^{*}\}$ can be covered by a finite number of balls of
radius arbitrarily close to $\veps$.
\end{proof}
Doeblin's condition which follows, is a criterion for
quasi-compactness (it should not be confused with its stronger
version, defined in Section \ref{ewak}, which was used in
\cite{sontag}).
\begin{definition}
\label{D-cond} Let $P(x,A)$ be a stochastic kernel defined on
$\mesp.$ We say that it satisfies Condition $D$ if there exist
$\theta>0, \eta<1$ and a probability measure $\mu$ on
$(\Omega,\mathcal B)$ such that
$$\mu (A)\geq \theta \Ra P(x,A) \geq \eta ~\mbox{for all}~x \in \Omega.$$
\end{definition}
\begin{example} \cite{doobs}\label{reason} 
  Condition D holds if $P(x,A) \leq K\mu(A)$ for some $K>0$ and
  probability measure $\mu \in \mathcal E$ (e.g.,
  $P(x,A)=\int_{A}p(x,y)\mu(dy)$ and $|p(x,y)| <K$).
\end{example}
\begin{theorem}
\label{main} Let $\mathcal M$ be an MCS. If $\Sigma$ is finite and 
for some $n \in \nn,$ all stochastic kernels $P_w(x,A),~w \in
\Sigma^n,$ satisfy Condition D, then $\calm$ is quasi-compact.
\end{theorem}
The proof, given in Appendix \ref{main-proof}, follows
the proof in \cite{yosida} that Condition D implies
quasi-compactness for an {\em individual} Markov operator.

The following lemma, whose proof is deferred to Appendix
\ref{proof-compact}, gives a complete characterization of a
quasi-compact MCS in terms of its associated Markov operators.
\begin{lemma}
\label{compact} If an MCS $\calm$ is quasi-compact, then 
$\alpha(T^*)=0,$ where $T^*=\{P_w: w \in \Sigma^*\}.$
\end{lemma}
It is easy to see that $\alpha(\calo)<\sup_{u \in \Sigma}
\alpha(P_u\calp) +\alpha(T),$ where $T=\{P_u: u \in \Sigma\}.$ This
yields a criterion for quasi-compactness in terms of the associated
Markov system $T$ and also suggests generalizations to infinite
alphabets, e.g. in the case if $\Sigma$ is a compact set and the map
$P(u)=P_u: \Sigma \ra \call$ is continuous.
\section{Weakly Ergodic MCS}
\label{ewak} 
For any Markov operator $P$ define \beq\label{wekly-erg}
\delta(P):= \sup_{\mu,\nu \in \calp}\half
\|P\mu-P\nu\|_1=\sup_{x,y}\sup_{A \in \calb}|P(x,A)-P(y,A)|. \feq Then
(we refer to \cite{iosifesku,iosif-theodor} for the properties of
Dobrushin's coefficient $\delta(P)$): \beqn \label{nn1}
\delta(P)=\sup_{ \lambda \in \caln \backslash \{0\}} \frac{\|P \lambda
  \|_1} {\|\lambda \|_1}, \feqn where $\caln=\{\lambda \in \cale:
\lambda(\Omega)=0\}.$
\begin{definition}
\label{weakerg} A Markov system $\{P_{u},~ u \in \Sigma\}$ is
called {\em weakly ergodic} if there exist constants $r,\delta >0$
such that $\delta(P_w) \leq 1-\delta$ for any $w \in \Sigma^r.$ An MCS
${\cal M}$ is called {\em weakly ergodic} if its associated Markov
system $\{P_{u}, ~u \in \Sigma\}$ is weakly ergodic.
\end{definition}
It follows from the definition and \eqref{nn1} that $\delta(P_w) \leq
M(1-\delta)^{|w|/r},$ for any $w \in \Sigma^*$ and some $M>0.$ Maass
and Sontag used a strong Doeblin's condition to prove the
computational power of noisy neural networks \cite{sontag}. They
essentially proved (see also \cite{paz,rabin}) the following result:
\begin{theorem}
\label{def} Let ${\cal M}$ be a weakly ergodic MCS. If a language
$L$ can be recognized by $\cal M$, then it is definite.
\end{theorem}
\begin{definition}
\label{doeblin} A Markov operator $P$ satisfies Condition $D_0$ if
$P(x,\cdot) \geq c \varphi (\cdot)$ for some constant $c \in
(0,1)$ and a probability measure $\varphi \in \calp$.
\end{definition}
If a Markov operator $P$ satisfies Condition $D_{0}$ with a
constant c, then  $\delta(P) \leq 1-c$ \cite{doobs}. The following
example shows that this condition is not necessary.
\begin{example} 
  Let $\Omega=\{1,2,3\}$ and $P(x,y)=\half$ if $x \neq y.$ Then
  $\delta(P)=\half$, but $P$ does not satisfy condition $D_0$.
\end{example}
We next state a general version of the Rabin-Paz stability theorem
\cite{paz,rabin}. We first define two MCS, $\calm$ and $\witi \calm$
to be {\em similar} if they share the same measurable space
$\mesp$, alphabet $\Sigma$, and sets $\cal A$ and $\cal R$,
and differ only in their Markov operators.
\begin{theorem}
\label{main-stab} Let ${\cal M}$ and $\witi{\cal M}$ be two
similar MCS such that the first is weakly ergodic. Then there is
$\alpha>0$, such that if $\|P_{u}-\tilde{P}_{u}\|_{1} \leq \alpha$
for all $u \in \Sigma$, then the second is also weakly ergodic.
Moreover, the two MCS recognize the same language.
\end{theorem}
For the sake of completeness we give a proof in Appendix \ref{paz-proof}.
\appendix \section*{Appendices}
\section{Proof of Theorem \ref{main}}
\label{main-proof}
\begin{lemma}\cite{yosida}
\label{we-com} Let $K(x,A)$ and $N(x,A)$ be two stochastic kernels
defined by
$$K(x,A)=\int_{A}k(x,y)\mu(dx),~~|k(x,y)| \leq C_K,$$
$$N(x,A)=\int_{A}n(x,y)\mu(dx),~~|n(x,y)| \leq C_N,$$
where $k(x,y)$ and $n(x,y)$ are measurable and bounded functions
in $\Omega \times \Omega,$ and $C_K,C_N$ are constants. Then $NK
\in \mathcal L$ is compact.
\end{lemma}
The proof in \cite{yosida} is for a special case, so we give here
an alternative proof.
\begin{proof}
Let $\{n_m(x,y):m \in \nn\}$ be a set of simple
and measurable functions such that \beq
\int_{\Omega}\int_{\Omega}|n_m(x,y)-n(x,y) 
|\mu(dx)\mu(dy) \leq
\frac{1}{m},\feq and define stochastic kernels $N_m(x,A)=\int_A
n_m(x,y)\mu(dy)$. Since the corresponding operators $N_{m} \in
\mathcal L$ have finite dimensional ranges they are compact. On
the other hand \beq \|NK-N_mK\|_{1}=\sup_{\|\varphi\|_{1} =1}
\|NK\varphi -N_{m}K\varphi \|_{1} \leq C_K/m, \feq thus,
$NK=\lim_{m \ra \infty}N_{m}K$ is a compact operator.
\end{proof}
Since operators $P_u,~u \in \Sigma$ satisfy Condition D, they can be
represented as $P_u=Q_u+R_u$, where $Q_u$ is defined by a stochastic
kernels having bounded and measurable on $\Omega \times \Omega$
densities $q_{u}(x,y)$ with respect to $\mu$, and $\|R_{u}\|_{1} \leq
1 -\eta$ \cite{yosida}. Consider the expansion of
$P_{w}=\prod_{k=0}^{m}(Q_{w_{k}}+R _{w_{k}}),~w \in \Sigma^{m+1}$ in
$2^{m+1}$ terms: \beq P_w=\prod_{k=0}^m Q_{w_k}+\sum_{j=0}^m \left(
  \prod_{k=1}^{j-1}Q_{w_k} R_{w_j} \prod_{k=j+1}^mQ_{w_k} \right)+
\ldots +\prod_{k=0}^mR_{w_k}.\feq By Lemma \ref{we-com}, the terms
contains $Q_{w_{i}}$ at least twice as factor are all compact
operators in $\mathcal L$. Since there are at most $m+2$ terms where
$Q_{w_{i}}$ appear at most once, then we obtain that for any $w \in
\Sigma^{m+1}$ there is a compact operator $Q_{w}$ such that
$\|P_w-Q_w\|_1 \leq (m+2) \cdot (1-\eta)^m. $
\section{Proof of Lemma \ref{compact}}
\label{proof-compact} 
We need the following proposition suggested to us by Leonid Gurvits.
\begin{proposition}
\label{gur} Let $Q_{1},Q_{2} \in \cal L$ be two compact operators,
and let $H=\{P_j\}\subseteq {\cal L}$ be a bounded set of
operators. Then, the set $Q=\{Q_{2}P Q_{1}: P \in H\}$ is totally
bounded.
\end{proposition}
\begin{proof}
  Let ${\cal K}= \{\mu \in \cale: \|\mu\|_1 \leq 1 \}$ and $X_i
  \subseteq {\cal E}: i=1,2$ be two compact sets such that $Q_i{\cal
    K} \subseteq X_i$. Define a bounded family ${\cal F}=\{f_{j}\}$ of
  continuous linear functions from $X_{1}$ to $X_{2}$ by setting
  $f_{j}=Q_{2}P_{j}$. Since $H$ is bounded, then ${\cal F} \subseteq C
  \left (X_{1},X_{2} \right )$ is bounded and equicontinuous, that is
  by Ascoli's theorem it is conditionally compact. Fix any $\veps>0$
  and consider a finite covering of $\calf$ by balls with radii
  $\veps.$ If $f_i$ and $f_j$ are included in the same ball, then
\begin{eqnarray*}
\|Q_2P_iQ_1-Q_2P_jQ_1\|_1 \leq \sup_{x \in
X_1}\|f_i(x)-f_j(x)\|_1\leq 2\veps.
\end{eqnarray*}
Therefore $\alpha(Q) \leq 2\veps$. This completes the proof since
$\veps$ is arbitrary.
\end{proof}
From Proposition \ref{gur} it follows that the set $\{Q_uPQ_v: u,v
\in \Sigma^n, P \in \call, \|P\|_1=1\}$ is totally bounded.

Fix any $\veps>0.$ There exist a number $n \in \nn$ and compact
operators $Q_w, ~w \in \Sigma^n$ such that $\|P_w-Q_w\|_1 \leq
\veps$ for all $w \in \Sigma^n.$ Since any word $w \in
\Sigma^{\geq 2n+1}$ can be represented in the form $w=u \hat w v$,
where $u,v \in \Sigma^n,$ and
\begin{eqnarray*}
\|P_{w}- Q_{v}P_{\hat {w}}Q_{u}\|_{1}&=&\|P_{v}P_{\hat {w}}P_{u}-
Q_{v}P_{\hat {w}}Q_{u}\|_{1} \leq \\
&\leq& \|P_{v}P_{\hat {w}}P_{u}- P_{v}P_{\hat {w}}Q_{u}\|_{1} +
\| P_{v}P_{\hat {w}}Q_{u}-Q_{v}P_{\hat {w}}Q_{u}\|_{1} \leq \\
&\leq& \|P_{u}-Q_{u}\|_{1} +\| P_{v}-Q_{v}\|_{1} \leq 2 \veps,
\end{eqnarray*}
we can conclude that $\alpha(T^{\geq 2n+1}) \leq 2\veps$, where
$T^{\geq 2n+1}=\{P_{w}: w \in \Sigma^{ \geq 2n+1}\}$. It follows
that $\alpha(T^{*})= \alpha(T^{\geq 2n+1}) \leq 2\veps,$
completing the proof since $\veps>0$ is arbitrary.
\section{Proof of Theorem \ref{main-stab}}
\label{paz-proof}
This result is implied by the following lemma:
\begin{lemma}
\label{paz} Let $\calm$ and $\witi \calm$ be two similar MCS, such
that the first is weakly ergodic and the second is arbitrary.  Then,
for any $\beta >0$ there exists $\veps >0$ such that
$\|P_u-\tilde{P}_u\|_1 \leq \veps$ for all $u \in \Sigma$ implies
$\|P_w-\tilde P_w\|_1 \leq \beta$ for all words $w \in
\Sigma^*$.
\end{lemma}
\begin{proof}
It is easy verify by using the representation \eqref{nn1} that:
\begin{itemize}
\item [(i)] For any Markov operators $P,Q,$ and $R,$ we have
$\|PQ-PR\|_1 \leq \delta(P)\|Q-R\|_1.$ \item[(ii)] For any Markov
operators $P,\tilde P$ we have $\delta(\tilde P)\leq
\delta(P)+\|P-\tilde P\|_1.$
\end{itemize}
Let $r \in \nn$ be such that $\delta(P_w) \leq \beta/7$ for any $w
\in \Sigma^r,$ and let $\veps=\beta/r.$ If $\|P_u-\tilde P_u\|_1
\leq \veps$ for any $u \in \Sigma,$ then  $\|P_w-\tilde P_w\|_1
\leq n \veps$ for any $w \in \Sigma^n.$ It follows that
$\|P_w-\tilde P_w\|_1 \leq \beta$ for any $w \in \Sigma^{\leq r}.$
Moreover, for any $v \in \Sigma^r$ and $w \in \Sigma^*,$ we have
\beq &&\|P_{vw}-\tilde P_{vw}\|_1 \leq
\|P_{vw}-P_v\|_1+\|P_v-\tilde P_v\|_1+\|\tilde P_v-\tilde
P_{vw}\|_1 \leq \nonumber\\&&\leq 2\delta(P_v)+\|P_v-\tilde
P_v\|_1+2\delta(\tilde P_v) \leq 4\delta(P_v)+3\|P_v-\tilde
P_v\|_1 \leq \beta,\feq completing the proof.
\end{proof}
\appendix
\section*{Acknowledgments}
\label{aknow} We are grateful to Leonid Gurvits for valuable 
discussions.

\end{document}